\documentclass[sigconf]{acmart}
\acmConference[EASE 2026]{The 30th International Conference on Evaluation and Assessment in Software Engineering}{9–12 June, 2026}{Glasgow, Scotland, United Kingdom}
\AtBeginDocument{%
  }

\setcopyright{acmlicensed}
\copyrightyear{2026}
\acmYear{2026}
\acmDOI{XXXXXXX.XXXXXXX}
\acmConference[EASE 2026]{The 30th International Conference on Evaluation and Assessment in Software Engineering}{9–12 June, 2026}{Glasgow, Scotland, United Kingdom}
\acmISBN{XXXXX}

\usepackage{comment}
\usepackage{enumitem}
\usepackage{makecell}
\usepackage{listings} 
\usepackage{fontawesome5}
\usepackage{pifont}
\usepackage{adjustbox}
\usepackage{booktabs}
\usepackage{caption}
\usepackage{subcaption}
\usepackage{array}
\usepackage{wrapfig}
\usepackage{algorithm}
\usepackage{multirow}
\usepackage{enumitem}
\usepackage{algorithmic}
\usepackage{hyperref}
\usepackage{tikz}
\usepackage{color}
\usepackage{pgfplots}
\usepackage[table]{xcolor}
\usepackage{textcomp}
\usepackage{graphicx}
\usepackage{algorithmic}
\usepackage[most]{tcolorbox}

\DeclareCaptionType{quoteFloat}[Quote][List of Quotes]
\DeclareRobustCommand{\ttt}[1]{{\small\texttt{\detokenize{#1}}}}

\newcommand{\docscite}[1]{\hyperlink{Tab:Sources}{[#1]}}

\colorlet{shadecolor}{gray!10}
\begin{document}

\title{Cache-Related Smells in GitLab CI/CD: Comprehensive Catalog, Automated Detection, and Empirical Evidence}


%% Each author must be defined separately for accurate metadata identification. As an exception, multiple authors may share one affiliation.

\author{Francesco Urdih}
\email{francesco.urdih@univie.ac.at}
\orcid{0009-0000-3507-5043}
\affiliation{%
  \institution{Software Architecture Research Group, Faculty of Computer Science, University of Vienna \and Doctoral School Computer Science DoCS, Faculty of Computer Science, University of Vienna}
  \city{Vienna}
  \country{Austria}
}

\author{Theodoros Theodoropoulos}
\email{theodoros.theodoropoulos@univie.ac.at}
\orcid{0000-0002-4618-4891}
\affiliation{%
  \institution{Software Architecture Research Group, Faculty of Computer Science, University of Vienna}
  \city{Vienna}
  \country{Austria}
}

\author{Uwe Zdun}
\email{uwe.zdun@univie.ac.at}
\orcid{0000-0002-6233-2591}
\affiliation{%
  \institution{Software Architecture Research Group, Faculty of Computer Science, University of Vienna}
  \city{Vienna}
  \country{Austria}
}

\renewcommand{\shorttitle}{Cache-Related Smells in GL CI/CD: Comprehensive Catalog, Automated Detection, and Empirical Evidence}
\renewcommand{\shortauthors}{Urdih et al.}

\begin{abstract}
Continuous Integration and Deployment (CI/CD) facilitate rapid software delivery, making fast feedback and minimal downtime essential. While caching has been shown to be an effective technique for tackling pipeline performance and reliability issues, existing works have primarily focused on missing dependency caches, ignoring other types of caches and cache misconfigurations. In this paper, we present a comprehensive catalog of ten cache-related smells in GitLab CI/CD that negatively impact performance and reliability, validated on a corpus of grey literature. To address the smells, we propose CROSSER, a tool that automatically detects seven of the ten smells. We evaluate CROSSER on a manually labeled dataset of 82 mature projects, achieving an overall $F_{1}$ score of 0.98. Finally, we investigate the presence of smells across a large dataset of 228 mature open-source projects and outline our empirical findings. Our results show a widespread frequency of the smells, as only 11\% of the projects do not present any. We also show that developers may not be aware of higher-level caching functionalities.
\end{abstract}


\keywords{CI/CD, Cache, Smells, Static Detection, Speed, Efficiency, Reliability}

\maketitle

\section{Introduction}
Continuous Integration and Continuous Deployment (CI/CD) are two software development practices to automate builds, tests, and deployments of software \cite{duvall2007continuous, shahin2017continuous}. Their adoption improves software quality \cite{hilton2016usage, chen2015continuous} and increases developers' productivity \cite{vasilescu2015quality, staahl2013experienced}.

Nevertheless, poorly configured CI/CD pipelines can lead to higher feedback times and higher resource consumption \cite{zhang2022buildsonic, gallaba2018use, vassallo2019automated, ghaleb2019empirical, bouzenia2024resource}, creating a barrier for the use of CI/CD \cite{widder2019conceptual, hilton2017trade}. To counter this, multiple works have proposed tools such as \ttt{CI-SKIPPER} \cite{abdalkareem2019commits} to determine whether a commit should trigger a pipeline, \ttt{HANSEL} \cite{gallaba2018use} to detect four CI misconfigurations, and \ttt{CD-LINTER} \cite{vassallo2020configuration} to find four CD smells. Furthermore, other studies have investigated what characterizes slow \cite{ghaleb2019empirical, vassallo2019automated, ghaleb2022studying} and failed \cite{ni2017cost, rausch2017empirical, ghaleb2022studying, vassallo2019automated} pipelines. The findings of Ghaleb et al.\ \cite{ghaleb2019empirical} show a strong association between fast pipelines and the use of caches. In a following study \cite{ghaleb2022studying}, they investigated both speed and reliability using a large dataset of 588 projects. The researchers indicate that caching is the most effective method for speeding up a workflow and reducing the number of broken pipelines. In addition, they manually investigated the repositories where caches did not provide notable pipeline improvements and identified the presence of misconfigurations.

In this work, we start from the findings of Ghaleb et al.\ and systematically study and detect cache-related issues that affect performance and reliability in CI/CD pipelines. Prior works \cite{gallaba2020accelerating, zhang2022buildsonic, celik2016build, esfahani2016cloudbuild} have already proposed tools to detect missing caches. However, they have mainly focused on dependency caches, overlooking other caches that can be used in CI/CD pipelines. Furthermore, CI/CD platforms include many functionalities related to caching. Such functionalities can lead to misconfigured caches, limiting their benefits and ultimately slowing pipelines or increasing the likelihood of failures. These aspects have not yet been studied or detected in other works. To help practitioners improve their pipeline configurations, we propose a novel reporting tool named \ttt{CROSSER}: {\em Cache-Related Observer of Smells for Speed, Efficiency, and Reliability}.

We focus our investigation on {\em GitLab CI/CD} given its popularity, especially among large companies \cite{gitlabUsage}, similarly to what other studies in the CI/CD community have done \cite{vassallo2020configuration, fairbanks2023analyzing, olewicki2022towards}. Nevertheless, we have investigated the presence of our proposed smells in another popular CI/CD tool, i.e., GitHub Actions. More specifically, we answer the following research questions:
\begin{itemize}[noitemsep,topsep=0pt]
    \item[\textbf{RQ1:}] What are cache-related smells relevant to practitioners that affect speed, efficiency, and reliability in GitLab CI/CD?
    \item[\textbf{RQ2:}] How accurately can \ttt{CROSSER} detect the smells identified in RQ1?
    \item[\textbf{RQ3:}] How frequent are the detected smells in mature open-source projects?
\end{itemize}

The contributions of our work are: \textbf{(i)} a catalog of ten cache-related smells relevant to practitioners affecting speed, efficiency, and reliability; \textbf{(ii)} \ttt{CROSSER}, an open-source tool covering seven of the ten smells, reaching an overall $F_{1}$ score of 0.98 on a dataset of 82 manually labeled mature open-source projects; \textbf{(iii)} an investigation of the frequency of the detected smells conducted on a dataset of 228 mature open-source projects, revealing a widespread presence of the smells, with a median of 3 smells per repository. Additionally, we add frequency-related considerations for two of the three smells not detected by \ttt{CROSSER}. All artifacts produced in this study are available in the replication package \cite{replicationPackage}. All the information needed to inspect these artifacts and run \ttt{CROSSER} is included. In addition, we included \ttt{CROSSER} in a GitLab repository\footnote{\url{https://gitlab.com/francesco.urdih/CROSSER/-/tree/version-paper}}.

This paper is organized as follows: Section \ref{Sec:Background} provides the background, while Section \ref{Sec:Methodology} describes the methodology. Section \ref{Sec:Results} illustrates the results, then discussed in Section \ref{Sec:Discussion}. Section \ref{Sec:Threats} examines potential validity threats that may influence our work. Section \ref{Sec:Related-Work} discusses related work, and, lastly, Section \ref{Sec:Conclusions} concludes our study.

\section{Background}
\label{Sec:Background}
In this section, we provide background information on CI/CD pipelines in GitLab. To understand how smells can occur, we focus on the available caching mechanisms.

\subsection*{GitLab CI/CD} Developers can configure CI/CD pipelines on GitLab by defining the \ttt{.gitlab-ci.yml} file in their repository. Each pipeline contains one or more {\em jobs}, which carry out {\em tasks} such as running unit tests, updating the documentation, etc. Each job belongs to a {\em stage}, with the default stages being \ttt{build}, \ttt{test}, and \ttt{deploy}. The jobs wait for all the jobs of the previous stages before running, unless developers change the execution graph with \ttt{needs}. Finally, developers can specify a Docker image which becomes the execution environment for one or more jobs, using the \ttt{image} keyword. An example of a workflow file is shown in Listing \ref{lst:workflow-example}. The workflow presents a build job, two test jobs, and a deployment job.

\subsection*{Caches in GitLab CI/CD} In GitLab, there are three ways to use caches. Two of them relate to raw files, and one to Docker. 

The first type of file cache is defined using the \ttt{artifacts} clause. This cache is accessible to jobs only when the pipeline is currently running. Afterward, artifacts can only be downloaded with the UI or with the REST API (e.g., to check logs). While the pipeline runs, the artifacts of a job are automatically fetched by following jobs unless configured otherwise. As an example, the \ttt{build_app} job from Listing \ref{lst:workflow-example} saves \ttt{app.jar}, which will be fetched and used by the testing jobs. The \ttt{deploy_app} job saves logs for later analysis.

The second type of file cache can be defined using the \ttt{cache} clause. Unlike \ttt{artifacts}, \ttt{cache} is designed to be used across multiple pipeline runs, and a job uses it only if explicitly configured. Any job pair can use the same \ttt{cache} by specifying a \ttt{key}. When sharing a key, jobs effectively work on the same files. For this reason, developers can specify the access type using \ttt{policy}. This clause accepts three values: \ttt{pull} for read-only access, \ttt{push} for write-only access, and \ttt{pull-push} (the default policy) for read and write access. As an example, the testing jobs from Listing \ref{lst:workflow-example} save and share the \ttt{gradle} and \ttt{coverage} files across pipeline runs. 

Depending on the functionality required, developers can choose the type of file cache to use. Typically, both types are used in the same workflow for different files, as shown in Listing \ref{lst:workflow-example}. To improve clarity, we will refer to the first cache type as {\em artifact}. Nevertheless, artifacts are effectively a file cache.

Lastly, GitLab workflows can use Docker caches. These are no different from the Docker caches employable on any device. For example, the \ttt{deploy_app} job from Listing \ref{lst:workflow-example} uses the layers from an old built image as build cache. However, GitLab presents some unique functionalities and limitations to Docker, as discussed in Section \ref{Sec:Smells}.

\definecolor{dark-green}{rgb}{0.0, 0.5, 0.0}

\begin{lstlisting}[language=make,
    basicstyle=\footnotesize\ttfamily,
    breakatwhitespace=true,         
    breaklines=true,                 
    captionpos=b,                    
    keepspaces=true,                 
    showspaces=false,                
    showstringspaces=false,
    showtabs=false,                  
    tabsize=2,
    emphstyle={\bfseries},
    emph={artifacts,cache,policy,key,paths},
    label=lst:workflow-example,
    caption={Example of a CI/CD pipeline to build, test, and deploy an application. The pipeline is affected by the smells presented in this work.}
]  
image: gradle:8

build_app:
  stage: build
  artifacts:
    paths: app.jar
  script: gradle jar

unit_testing:
  stage: unit-tests
  cache: # policy is pull-push by default
    key: code-coverage-$CI_COMMIT_REF_SLUG
    paths: 
      - .gradle/
      - .coverage/
  script: gradle test --tests 'unit.*' -PAppPath=/app.jar -PCoveragePath=/.coverage

integration_testing:
  stage: integration-tests
  cache: # policy is pull-push by default
    key: code-coverage-$CI_COMMIT_REF_SLUG
    paths: 
      - .gradle/
      - .coverage/
  script: gradle test --tests 'integration.*' -PAppPath=/app.jar -PCoveragePath=/.coverage

deploy_app:
  stage: deploy
  artifacts:
    paths: build.log
  script: 
    - docker build -t app/image . | tee build.log
    - docker push app/image
\end{lstlisting}

\section{Methodology}
\label{Sec:Methodology}

\begin{figure*}[htbp]
    \caption{The methodology applied in this work.} 
    \centerline{\includegraphics[width=0.85\textwidth]{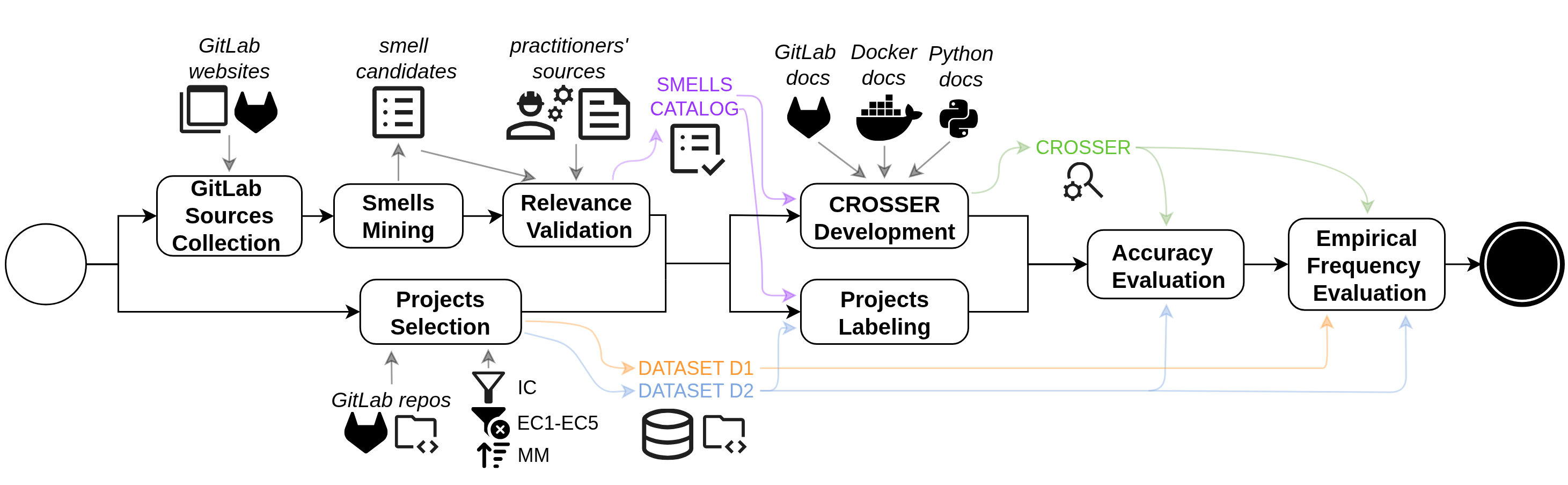}}
    \label{fig:methodology}
\end{figure*}

This section presents our research method, summarized in Fig.~\ref{fig:methodology}. 

\subsection{Smells Mining and Validation}
\label{Subsec:Smells-Catalog-Method}

\subsubsection*{Smells Mining}

Similarly to the approach of related studies \cite{gallaba2018use, zhang2022buildsonic, sharma2016does}, to produce an initial catalog of candidate smells, we analyzed all the official sources from the studied technology. In the case of GitLab CI/CD \cite{gitlabDocs}, the CI/CD documentation totaled $\sim$150 web pages. Furthermore, we analyzed other official GitLab webpages (e.g., blog posts, release notes) gathered with search engines (e.g., Bing, Google Search). Focusing on these official sources ensures that the identified smells accurately reflect the platform’s recommended practices and potential pitfalls. This enhances the catalog's credibility and relevance for real-world GitLab CI/CD usage.

The smells we identified are derived from the information of 27 sources (summarized in Table \ref{Tab:Sources}). Throughout this work, we often cite these sources with their ID (e.g., \docscite{S1}, \docscite{S2}) to support our claims. We followed an open coding procedure \cite{stol2016grounded} to analyze the sources and find explicit cache-related performance and reliability smells mentioned. Whenever possible, we re-formulated identified best practices as smells. For example, {\em God Cache} (see Section \ref{Sec:Smells}) is derived from the best practice {\em Use Multiple Caches} \docscite{S10}. 

All the smells we identified are present in GitLab version 18, i.e., the version available at the time of writing.

\hypertarget{Tab:Sources}{}
\begin{table}[htbp]
\caption{GitLab sources mentioning the smells catalog.}
\scriptsize
\begin{center}
\begin{tabular}{|c|l|l|}
\hline
\textbf{ID} & \makecell{\textbf{Title}} & \makecell{\textbf{Short Url}} \\
\hline \hline
S1 & CI/CD YAML syntax reference & https://tinyurl.com/8es4sje3 \\ \hline
S2 & Dependency proxy for container images & https://tinyurl.com/3hdnd723 \\ \hline
S3 & Using the Dependency Proxy to improve pipelines & https://tinyurl.com/yn4hwzzr \\ \hline
S4 & Known issues with Docker-in-Docker & https://tinyurl.com/mr2atkzp \\ \hline
S5 & Make Docker builds faster with layer caching & https://tinyurl.com/2vhyumxy \\ \hline
S6 & CI/CD Caching examples & https://tinyurl.com/abdm6em \\ \hline
S7 & Job Artifacts & https://tinyurl.com/msfbakpm \\ \hline
S8 & Artifact and cache settings & https://tinyurl.com/2pwb92h8 \\ \hline
S9 & GitLab-hosted runners & https://tinyurl.com/3f7esc3c \\ \hline
S10 & Caching in GitLab CI/CD & https://tinyurl.com/2ntksztb \\ \hline
S11 & Predefined CI/CD variables reference & https://tinyurl.com/4z5kumpe \\ \hline
S12 & Run your CI/CD jobs in Docker containers & https://tinyurl.com/mr4urwrk \\ \hline
S13 & Per-cache fallback cache keys in CI/CD pipelines & https://tinyurl.com/2tbs6r3a \\ \hline
S14 & Pipeline Efficiency & https://tinyurl.com/y5euxupf \\ \hline
S15 & Build Docker images with BuildKit & https://tinyurl.com/yc2j9jry \\ \hline
S16 & Troubleshooting Code Quality & https://tinyurl.com/vf6wz5et \\ \hline
S17 & Unit test reports & https://tinyurl.com/mryf3kmr \\ \hline
S18 & Migrating from GitHub Actions & https://tinyurl.com/4a4ec4vj \\ \hline
S19 & Reduce container registry data transfers & https://tinyurl.com/3nfpernz \\ \hline
S20 & Include report artifacts as release evidence & https://tinyurl.com/54sc3b32 \\ \hline
S21 & Create a GitLab Pages website from scratch & https://tinyurl.com/5n89xypv \\ \hline
S22 & Build and push container images to the registry & https://tinyurl.com/mr3ax96e \\ \hline
S23 & Make jobs start earlier with needs & https://tinyurl.com/3fkf8h8k \\ \hline
S24 & Dependency Proxy & https://tinyurl.com/2s48yu3s \\ \hline
S25 & Speed up job execution & https://tinyurl.com/ay9c7fm2 \\ \hline
S26 & Use multiple caches in the same job & https://tinyurl.com/mv325a67 \\ \hline
S27 & Set default artifacts expiration & https://tinyurl.com/ycyvxf69 \\ \hline
\end{tabular}
\label{Tab:Sources}
\end{center}
\end{table}  

\subsubsection*{Relevance Validation}

To assess whether the insights from GitLab's official materials align with practitioners' needs and experiences, we validated them against a curated body of practitioner literature (grey literature). Specifically, we relied on a pre-compiled, quality-checked corpus published by Urdih et al.~\cite{urdih2025architectural} that was compiled according to the guidelines of Rainer et al.~\cite{rainer2019using}. Grey literature offers direct insight into the concerns of practitioners and the state of practice, thus complementing official documentation and academic work \cite{garousi2019guidelines}. Our use of such sources is consistent with previous studies in which grey literature was used to create catalogs of patterns, best practices, anti-patterns, and smells \cite{zampetti2020empirical,warnett2022architectural,ricca2025multi}. 

This corpus contains sources discussing techniques to improve the performance and reliability of CI/CD pipelines, and in general, their quality. These encompass multiple CI/CD technologies (e.g., GitLab CI/CD, GitHub Actions) and offer a cross-tool perspective that enhances the generalizability of our observations. We therefore believe that the corpus is suitable for validating practitioners' interest in the mined smells. From the original 38 sources compiled by Urdih et al., we excluded 10 written by official CI/CD vendors to focus on practitioners' voices, leaving 28 blog-like sources. 

Two authors independently analyzed the 28 sources using an open coding procedure \cite{stol2016grounded}, and subsequently compared the resulting notes. Specifically, we read each source to extract text passages dealing with caching, performance, reliability, and related operational challenges and practices in CI/CD. We then coded these sections and compared the resulting codes with our preliminary smell catalog derived from GitLab sources, assigning practitioners' insights to specific smells where appropriate. Two illustrative examples of sources and their discussions are presented in Section \ref{Sec:Smells}. The complete set of analyzed sources and our coding notes are available in the replication package \cite{replicationPackage}. At the end of this phase, we retained only the candidate smells discussed by practitioners and removed the others.

\subsection{Projects Selection and Labeling}
\label{Subsec:Datasets}

In this section, we present the methods used to construct a dataset of 228 mature open-source GitLab projects (referred to as {\em D1}) to address RQ3. To answer RQ2, we further refine {\em D1} to a subset of 82 projects (denoted as {\em D2}). Since our selection process does not consider the repository's technical characteristics, {\em D1} includes projects with very different purposes and technologies, thereby increasing the reliability and generalizability of our results. In addition, we use inclusion/exclusion criteria to remove toy projects from our datasets. Please note that the mining took place in April 2025.

\subsubsection*{D1: Dataset for RQ3} We need a large and diverse set of CI/CD-related projects to determine how frequently the smells occur in practice. Thus, we mined a dataset of open-source systems from GitLab. We employ only popularity as \textbf{inclusion criterion} (IC): $100+$ stars. This filter is an efficient and effective method for finding mature candidates for software engineering investigations \cite{munaiah2017curating} and has been previously used in studies similar to our \cite{vassallo2020configuration, openja2022studying, zhang2019large, bouzenia2024resource}. 

After finding 400 repositories with our inclusion criterion, we employed the following \textbf{exclusion criteria} (EC):
\begin{itemize}[noitemsep]
    \item[\bf{EC1}] {\em Private files} (5 projects).
    \item[\bf{EC2}] {\em Missing CI/CD}: CI/CD files not included in the repository (135 projects).
    \item[\bf{EC3}] {\em Wrong CI/CD}: The workflow fails linting \cite{gitlabLint} (3 projects).
    \item[\bf{EC4}] {\em Insufficient usage of CI/CD}: Less than 10 successful pipeline runs in the entire repository's life (13 projects)\footnote{Notably, 37 repositories had pipeline runs accessible only by project members. We did not apply {\em EC4} for them.}.
    \item[\bf{EC5}] {\em Inactive}: No activity in the previous 18 months (13 projects)\footnote{{\em EC5} also ensures that our analysis does not include repositories too old to present some of the smells (i.e., the smell was introduced after the last commit).}.
    \item[\bf{EC6}] {\em Forks} of systems already in the dataset (3 projects).
\end{itemize}

At the end of this process, we were left with 228 projects. 

\subsubsection*{D2: Dataset for RQ2} Unlike RQ3, to answer RQ2, we needed a labeled dataset (denoted as {\em D2}). Due to the substantial manual effort required for labeling, we aimed to label 50 projects for each smell, as done by Zhang et al.\ \cite{zhang2022buildsonic}. We focused on the most mature projects within {\em D1}, as these are most likely to include complex CI/CD pipelines. We ranked the repositories in {\em D1} using five maturity metrics (MM) proposed by Munaiah et al.~\cite{munaiah2017curating}: number of core contributors, comment ratio, commit frequency, issue frequency, and software license. More details on the ranking process are included in the replication package \cite{replicationPackage}. Ultimately, we selected the 50 repositories with the highest rankings. Only five of the seven detected smells are always applicable (see Section \ref{Sec:Smells}); one is related only to groups, and another is detected by \ttt{CROSSER} only to pipelines with Python. Since 48 out of the 50 highest-ranked repositories are within groups, we added 2 more. Out of these 52, 20 use Python, so we added 30 more, finally obtaining the dataset {\em D2} with 82 projects. 

After completing the selection phase, two authors (one of whom did not contribute to the development of \ttt{CROSSER}) manually inspected the CI/CD workflow files and independently identified the seven smells that \ttt{CROSSER} detects. For the few disagreements encountered, the third author helped in assigning the label. 

\subsection{CROSSER Development}
\label{Subsec:CROSSER}

This section provides an overview of \ttt{CROSSER} and its development process. \ttt{CROSSER} is a static analysis tool written in \ttt{Kotlin} that detects seven of the ten cache-related smells proposed in this study. \ttt{CROSSER} functions similarly to tools proposed by other researchers \cite{gallaba2018use, zhang2022buildsonic, vassallo2020configuration}: to identify smells in a repository, it parses the workflow file and maps the values to a meta-model. This meta-model contains only the information relevant for the detection of smells (e.g., the \ttt{cache}, \ttt{artifacts} clauses \docscite{S1}), and it ignores the irrelevant one (e.g., the \ttt{secrets}, \ttt{identity}, clauses \docscite{S1}). Once the parsing phase is completed, the smells are identified by checking specific meta-model values or aggregating multiple values across the workflow. 

The parsing component is essential for the accuracy of our tool. Given this, we rely on the \ttt{SnakeYAML} library \cite{snakeYAML}. After studying the syntax of GitLab workflow files \docscite{S1} in depth, we carefully selected and saved all the YAML keywords needed to identify the smells. Additionally, we have developed unit tests that verify the correctness of each parsed keyword in multiple example workflows. Furthermore, it is essential to mention that GitLab workflows contain several features that make the parsing phase not trivial, such as: global default values (\ttt{default} clause \docscite{S1}), variables reducing the usage of hardcoded values (\ttt{variables} clause \docscite{S1}), jobs inheritance to inherit values (\ttt{extends} clause \docscite{S1}), etc. Given this complexity, we have developed several integration tests to ensure that the extracted meta-models correctly represent the workflows, after the related component has resolved all inheritance logic. 

\subsection{Accuracy and Frequency Evaluation}

To evaluate the accuracy of \ttt{CROSSER}, we have run the tool on {\em D2} and processed the $F_1$ score. Then, we manually reviewed the YAML files for the false positives and false negatives to understand the causes of \ttt{CROSSER}'s misclassifications, drawing insights into its accuracy.

To evaluate the frequency of smells and obtain empirical findings, we followed a mixture of qualitative and quantitative analyses. We have run \ttt{CROSSER} on {\em D1} to extract frequency metrics, and incorporated examples encountered during the labeling of {\em D2}. Notably, we have included some considerations over the potential frequency of two of the three smells that \ttt{CROSSER} does not automatically detect.

\section{Results}
\label{Sec:Results}

In this section, we present the results for our research questions, summarized in Tables \ref{Tab:RQ1}, \ref{Tab:RQ2}, and \ref{Tab:RQ3-summary}. 

\subsection{Smells Catalog and Automatic Smell Detection}
\label{Sec:Smells}

In this segment, we address RQ1 by presenting the results of our smells mining and validation process. We have identified ten cache-related smells affecting speed, efficiency, and reliability in GitLab CI/CD pipelines, which are relevant to practitioners. The smells are summarized in Table \ref{Tab:RQ1}. In total, all smells affect efficiency, nine affect speed, and six affect reliability. Six smells are cache misconfigurations, while four can be described as a {\em missing cache}. To the best of our knowledge, only two of these smells have been automatically detected in related work (see Table \ref{Tab:RQ1}). This is despite the fact that seven out of the ten smells are also present in another CI/CD platform, i.e., GitHub Actions (GHA). The references to GHA are included in the replication package \cite{replicationPackage}.

As outlined in Section \ref{Subsec:Smells-Catalog-Method}, to answer RQ1, we first examined the official GitLab sources to create the smells candidates. Then, we studied practitioners' sources to understand the relevance of these candidates. After the first phase, the catalog contained 12 smells. Two of these are not mentioned in any practitioner source and were, therefore, removed. The most frequently discussed smell is {\em No Dependencies Cache}, which is the most common use of a cache and has been studied in detail by related work. Additionally, eight other smells are mentioned in two or more practitioners' sources. We present examples of relevant discussions from two practitioners' sources at the end of this section.

Before that, we present the ten validated smells and describe our proposed detection strategy as implemented in \ttt{CROSSER}, for seven of these smells. Three smells pose particular challenges that prevent automatic detection. 

\newcolumntype{U}[1]{>{\centering\arraybackslash}p{#1}}

\begin{table*}[htbp]
\caption{Summary of RQ1 results: the catalog of cache-related smells affecting speed, efficiency, and reliability that are discussed in practitioners' sources. SP = speed, EF = efficiency, RE = reliability. GHA = GitHub Actions.}
\begin{center}
\rowcolors{4}{gray!25}{white}
\begin{tabular}{|p{4.5cm}|U{2.2cm}|U{0.4cm}U{0.4cm}U{0.4cm}|U{2.05cm}|U{1.95cm}|U{1.4cm}|U{1.2cm}|}
\hline
\multirow{2}{*}{\centering\textbf{Smell}\arraybackslash} &
\multirow{2}{*}{\textbf{Smell Type}} 
& \multicolumn{3}{c|}{\textbf{Affecting}} 
& \multirow{2}{*}{\makecell{\textbf{Practitioner}\\\textbf{Sources Count}}}
& \multirow{2}{*}{\makecell{\textbf{Detected by}\\\textbf{Related Work}}} 
& \multirow{2}{*}{\makecell{\textbf{Detected}\\\textbf{by} \ttt{CROSSER}}} 
& \multirow{2}{*}{\textbf{In GHA}} \\
\cline{3-5}
& & \textbf{SP} & \textbf{EF} & \textbf{RE} & & & & \\
\hline
\hline
Artifacts Default Expiration Period & Misconfiguration & & \faCheck & \faCheck & 3 & & \faCheck & \faCheck \\ \hline
Artifacts Fetched by Default  & Misconfiguration   & \faCheck & \faCheck & & 3 & & \faCheck & \\ \hline
No Dependencies Cache & Missing Cache & \faCheck & \faCheck & \faCheck & 15 & \cite{zhang2022buildsonic,gallaba2020accelerating,esfahani2016cloudbuild,celik2016build} & \faCheck & \faCheck \\ \hline
No Fallback Cache & Missing Cache & \faCheck & \faCheck & \faCheck & 2 & & \faCheck & \faCheck \\ \hline
God Cache & Misconfiguration & \faCheck & \faCheck &  & 2 & & & \faCheck \\ \hline
Use of a Protected Cache & Misconfiguration & \faCheck & \faCheck & \faCheck & 2 & & & \\ \hline
Redundant Cache Updates & Misconfiguration  & \faCheck & \faCheck & & 2 & & \faCheck & \faCheck  \\ \hline
Suboptimal Compression Type & Misconfiguration & \faCheck & \faCheck & & 4 & & & \faCheck \\ \hline
No Docker Pull-Through Cache & Missing Cache & \faCheck & \faCheck & \faCheck & 4 & & \faCheck & \\ \hline
No Docker Layers Cache & Missing Cache & \faCheck & \faCheck & \faCheck & 4 & \cite{gallaba2020accelerating} & \faCheck & \faCheck \\

\hline
\end{tabular}
\label{Tab:RQ1}
\end{center}
\end{table*}

\subsubsection*{SM1: Artifacts Default Expiration Period}\mbox{}

{\bf Description}. GitLab artifacts are retained only for a certain period, set with \ttt{artifacts:expire_in}. If the value is not specified, \ttt{gitlab.com} saves them for 30 days \docscite{S27}. This extended retention period can lead to inefficient use of storage space, particularly for artifacts needed only while the pipeline runs \docscite{S1, S16}. Additionally, not setting the value may lead to reliability issues if the artifacts are needed for more than 30 days \docscite{S1, S7, S16, S17, S20}. 

%Besides efficiency and reliability, this smell removes one key functionality for artifacts: keeping specific artifacts after the expiration date if desired [S7]. This option is handy when developers want to maintain certain logs for future investigation. Nevertheless, the feature is not available when the expiration period is left as the default, forcing developers to use a secondary storage space in this scenario. 

{\bf Example}. The \ttt{build_app} job from Listing \ref{lst:workflow-example} defines artifacts without the \ttt{expire_in} clause. In this pipeline, \ttt{app.jar} is retained by GitLab for 30 days, even though it is needed only until the testing jobs start. Additionally, the \ttt{deploy_app} job saves the build logs for future analysis. These could be needed for a long time, but are retained for only 30 days. In Listing \ref{lst:workflow-repaired}, \ttt{app.jar} is retained only for 1 hour, while the logs for 3 months.

{\bf Detection Strategy}. \ttt{CROSSER} identifies this smell whenever artifacts are defined, but \ttt{expire_in} is not set.

\subsubsection*{SM2: Artifacts Fetched by Default}\mbox{}

{\bf Description}. When a job saves artifacts, all waiting jobs\footnote{\footnotesize\texttt{jobB} waits for \footnotesize\texttt{jobA} if it runs in one of the next stages, or if it declares an explicit dependency on \footnotesize\texttt{jobA} using \footnotesize\texttt{needs}.} will, by default, fetch them \docscite{S1}, regardless of whether they are actually useful. This default behavior can significantly impact the efficiency and performance of the pipeline, especially for large artifacts. To prevent it, developers must explicitly define artifact dependencies using the \ttt{dependencies} or \ttt{needs:artifacts} clauses. 

{\bf Example}. The job \ttt{deploy_app} from Listing \ref{lst:workflow-example} fetches the \ttt{app.jar} defined by the \ttt{build_app} job, even though it is not necessary for building the Docker image. This ultimately slows down the task. In Listing \ref{lst:workflow-repaired}, all artifact dependencies are made explicit.

{\bf Detection Strategy}. To detect this smell, \ttt{CROSSER} first builds the execution graph of the pipeline \docscite{S23} based on the \ttt{stage} and \ttt{needs} of each job. Using the graph, the smell is found for each pair, \ttt{jobA} and \ttt{jobB}, for which the following conditions are met: \textit{(1)} \ttt{jobA} defines artifacts, \textit{(2)} \ttt{jobB} waits for \ttt{jobA}, \textit{(3)} \ttt{jobB} does not specify its dependencies with the \ttt{dependencies} or \ttt{needs:artifacts} clauses.

\subsubsection*{SM3: No Dependencies Cache}\mbox{}

{\bf Description}. Almost every project relies on dependencies to work. These need to be downloaded every time and are rarely changed. Downloading dependencies at every run slows down the pipeline, increases inefficiencies, and increases the likelihood of a failing pipeline \docscite{S1, S6, S14, S18, S21}. Therefore, developers should use a cache for the jobs that install dependencies. 

{\bf Example}. The \ttt{build_app} job from Listing \ref{lst:workflow-example} installs Java libraries at every run, slowing down the build task. In Listing \ref{lst:workflow-repaired}, the installation takes place only when new libraries are added.

{\bf Detection Strategy}. To detect this smell, \ttt{CROSSER} must first detect whether a job is downloading dependencies. Given the numerous dependencies that any application can rely on, we focus on Python libraries due to their widespread use. We have studied the documentation of \ttt{pip} \cite{pipDocs} and \ttt{conda} \cite{condaDocs}, and we have identified the commands to download and install libraries, as well as the typical location where these are saved. Additionally, Linux-based systems can install Python libraries with package managers like \ttt{apt-get} \cite{scipyDocs}. With this information, \ttt{CROSSER} identifies the {\em No Dependencies Cache} smell by checking whether libraries are downloaded by the job (i.e., \ttt{before_script}, \ttt{script}, and \ttt{after_script} clauses) and whether the expected paths are listed in \ttt{cache:paths}.

\subsubsection*{SM4: No Fallback Cache}\mbox{}

{\bf Description}. The cost of a cache miss in terms of speed, efficiency, and reliability can be considerable. Therefore, jobs in GitLab should define secondary keys pointing to fallback caches using \ttt{cache:fallback_keys}, employed if the original key is not found \docscite{S1, S10, S13}. Furthermore, GitLab allows defining a global fallback key setting the \ttt{CACHE_FALLBACK_KEY} variable \docscite{S1, S10}.

{\bf Example}. Listing \ref{lst:workflow-example} uses a branch-based key, i.e., jobs from different branches use different caches. When a new branch is created, the job will start without cache. In the case of Listing \ref{lst:workflow-repaired}, the jobs use the default branch as a fallback cache, so they will start with the cache of the default branch and then push a new branch-specific cache, reducing the number of cold starts.

{\bf Detection Strategy}. When \ttt{cache} is used, \ttt{CROSSER} checks whether the global fallback key or the job-based keys are defined.

\subsubsection*{SM5: God Cache}\mbox{}

{\bf Description}. Like a {\em God Class} \cite{riel1996object} holds too many responsibilities, a {\em God Cache} saves too many files with different purposes. This slows down jobs that need only some of the files \docscite{S10, S26}. To avoid the smell, developers must split the paths across multiple keys in the \ttt{cache} clause.

{\bf Example}. The testing jobs from Listing \ref{lst:workflow-example} save all the files under a single key. In Listing \ref{lst:workflow-repaired}, the build and test coverage files are saved under two keys. Job \ttt{build_app} will only fetch the former. 

{\bf Detection Strategy}. As outlined for {\em No Dependencies Cache}, there are numerous tools that need caches. Even for a single tool, it is often good to create multiple caches. An example is separating development and production dependencies. For this reason, we do not detect this smell in \ttt{CROSSER}.

\subsubsection*{SM6: Use of a Protected Cache}\mbox{}

{\bf Description}. GitLab repositories save caches in two registries \docscite{S10}: one for protected branches (e.g., \ttt{master}) and one for unprotected branches. When a pipeline runs from an unprotected branch, jobs cannot fetch the cache of protected branches, even if the cache key would match. This behavior is enabled by default for security reasons, preventing developers with low-level privileges from accessing hidden files through the cache. If security is a concern, this functionality should not be disabled. In all other cases, using two caches increases disk usage and the likelihood of cache misses, worsening performance, efficiency, and reliability. 

{\bf Detection Strategy}. The project configuration is known only to users with a GitLab role in the said project. For this reason, we do not identify it in our tool.

\subsubsection*{SM7: Redundant Cache Updates}\mbox{}

{\bf Description}. GitLab allows many jobs with different purposes. While some of these jobs can reuse the same cached files, only one would need to update them to reflect the new commit. Multiple updates to the same cached files ultimately slow down the pipeline \docscite{S1, S6}. For this reason, the access strategy can be defined with \ttt{cache:policy} (see Section \ref{Sec:Background}). To avoid this smell, developers should either assign unique caches to jobs or ensure that no more than one job is configured with the \ttt{push} or \ttt{pull-push} policy.

{\bf Example}. The testing jobs from Listing \ref{lst:workflow-example} both update the same cache, even though \ttt{integration_testing} could just pull the cache.

{\bf Detection Strategy}. To detect this smell, \ttt{CROSSER} focuses entirely on the \ttt{cache} clause of all pipeline jobs. The smell is found for any pair of jobs, \ttt{jobA} and \ttt{jobB}, that meet the following two conditions: \textit{(1)} \ttt{jobA} and \ttt{jobB} use the same cache key, and this key does not contain variables unique to the job \docscite{S11}, such as {\small\texttt{\$CI\_JOB\_ID}}. \textit{(2)} \ttt{jobA} and \ttt{jobB} use either a \ttt{push} or a \ttt{pull-push} policy. 

Additionally, this smell may be found within matrix jobs\footnote{Matrix jobs are jobs declared as one in the configuration, but that result in multiple jobs when the pipeline runs. They are typically used to test multiple software versions in a concise way.} \docscite{S1} that push to the same cache.

\subsubsection*{SM8: Suboptimal Compression Type}\mbox{}

{\bf Description}. When a job starts, artifacts and caches must be downloaded and decompressed locally. Similarly, before a job ends, they must be compressed and uploaded to GitLab. GitLab has five levels of compression \docscite{S8}, which differ in the degree of file compression and, therefore, influence the time required for compression, decompression, upload, and download. Selecting an optimal compression level depends on the network speed, the type of files, and the runner's hardware, as compression time is hardware-dependent.

The default compression level prioritizes compression time over compressed file size. It may be suboptimal for speed, particularly if the job does not use GitLab's standard runners \docscite{S8, S9}. Additionally, it is not ideal for large files that should be more compressed.

{\bf Example}. The \ttt{deploy_app} job from Listing \ref{lst:workflow-example} saves build logs, which can impact storage, especially when many pipelines run daily. In Listing \ref{lst:workflow-repaired}, a higher compression is used.

{\bf Detection Strategy}. It is not feasible to automatically determine whether a repository uses the optimal compression level, as it depends on the types of files cached, the runners employed, and the specific developers' requirements (i.e., efficiency vs. speed).

\subsubsection*{SM9: No Docker Pull-Through Cache}\mbox{}

{\bf Description}. Docker images can be retrieved from container registries in GitLab pipelines by either specifying the \ttt{image} clause \docscite{S1} or executing commands such as \ttt{docker pull}. GitLab offers to group repositories \docscite{S2} a mirror to Docker Hub \cite{dockerMirror} that acts as a pull-through cache, saving images locally on the GitLab servers. This functionality enables faster image fetching, avoiding rate limits and service interruptions on Docker Hub \docscite{S2, S3, S22, S24, S25}. 

To use the cache, developers must specify the mirror registry when pulling images. This can be done directly with its identifier (\ttt{gitlab.com/group_name/dependency_proxy}) or with a predefined variable \docscite{S11}, such as {\small\texttt{\$DEPENDENCY\_PROXY}}.

{\bf Example}. All jobs in Listing \ref{lst:workflow-example} pull the \ttt{gradle:8} image directly from Docker Hub, which slows down the jobs and makes them vulnerable to the platform's rate limits and malfunctions. In Listing \ref{lst:workflow-repaired}, the pull goes through the GitLab dependency proxy instead.

{\bf Detection Strategy}. First, \ttt{CROSSER} checks whether a repository is part of a group (the information was originally obtained from the GitLab API). Then, for group repositories, it extracts the container registry of each image pulled and checks if any of these are pulled directly from Docker Hub. If the registry is omitted in the image identifier, GitLab defaults to Docker Hub unless the job has been authenticated with a different registry.

To extract the registry, \ttt{CROSSER} attempts to split the image identifier, typically written as \ttt{registry/image-name:tag-name}. 

\subsubsection*{SM10: No Docker Layers Cache}\mbox{}

{\bf Description}. When building Docker images, Docker attempts to use previously built layers to accelerate the build process. However, each GitLab job starts with an empty Docker Daemon \docscite{S4}, preventing the reuse of cached layers by default. To overcome this limitation, users can manually specify an existing image whose layers should be used as a cache. This can be done with the {\small\texttt{-{}-cache-from}} option of the \ttt{docker build} command \docscite{S5, S15, S19}. Omitting this option results in the image being built from scratch. 

{\bf Example}. The \ttt{deploy_app} job from Listing \ref{lst:workflow-example} builds an image without using a layers cache, while Listing \ref{lst:workflow-repaired} uses it.

{\bf Detection Strategy}. \ttt{CROSSER} checks each job's steps for \ttt{docker build} (and equivalent commands) with missing {\small\texttt{-{}-cache-from}}.

\begin{lstlisting}[language=make,
    basicstyle=\footnotesize\ttfamily,
    breakatwhitespace=true,         
    breaklines=true,                 
    captionpos=b,                    
    keepspaces=true,                 
    showspaces=false,                
    showstringspaces=false,
    showtabs=false, 
    alsoletter={-},
    emph={dependencies, expire_in, $CI_DEPENDENCY_PROXY_GROUP_IMAGE_PREFIX, policy, --cache-from, ARTIFACT_COMPRESSION_LEVEL,fallback_keys},
    emphstyle={\color{dark-green}},
    label=lst:workflow-repaired,
    caption={The workflow from Listing \ref{lst:workflow-example}, repaired. }
]  
image: $CI_DEPENDENCY_PROXY_GROUP_IMAGE_PREFIX/gradle:8

build_app:
  stage: build
  artifacts:
    paths: app.jar
    expire_in: 1 hour
  cache:
    key: build-$CI_COMMIT_REF_SLUG
    paths: .gradle
    fallback_keys: [build-$CI_DEFAULT_BRANCH]
  script: gradle jar

unit_testing:
  stage: unit-tests
  dependencies: [build_app]
  cache: 
    - key: build-$CI_COMMIT_REF_SLUG
      paths: .gradle
      policy: pull
      fallback_keys: [build-$CI_DEFAULT_BRANCH]
    - key: code-coverage-$CI_COMMIT_REF_SLUG
      paths: .coverage/
      fallback_keys: [code-coverage-$CI_DEFAULT_BRANCH]
  script: gradle test --tests 'unit.*' -PAppPath=/app.jar -PCoveragePath=/.coverage

integration_testing:
  stage: integration-tests
  dependencies: [build_app]
  cache: 
    - key: build-$CI_COMMIT_REF_SLUG
      paths: .gradle
      policy: pull
      fallback_keys: [build-$CI_DEFAULT_BRANCH]
    - key: code-coverage-$CI_COMMIT_REF_SLUG
      paths: .coverage/
      policy: pull
      fallback_keys: [code-coverage-$CI_DEFAULT_BRANCH]
  script: gradle test --tests 'integration.*' -PAppPath=/app.jar -PCoveragePath=/.coverage

deploy_app:
  stage: deploy
  dependencies: []
  artifacts:
    paths: build.log
    expire_in: 3 months
  variables:
    FF_USE_FASTZIP: 1
    ARTIFACT_COMPRESSION_LEVEL: slowest
  script: 
    - docker pull app/image || true
    - docker build --cache-from app/image -t app/image . | tee build.log
    - docker push app/image
\end{lstlisting}

%\subsubsection*{Presence in GitHub Actions}\mbox{}

%As shown in the last column of Table \ref{Tab:RQ1}, six of the nine proposed smells are also present in GitHub Actions. Both artifacts and caches are available for GitHub workflows, and they work in a similar fashion. Expiration periods can be defined for artifacts, although they can be left as a default. Similarly, the compression level can be defined for artifacts (we did not find evidence of the compression level for caches). Secondly, dependencies can be installed without using a cache, and too many files may be saved in a cache, leading to a {\em God Cache}. In addition, caches can be configured to be downloaded or uploaded only. Lastly, when building a Docker image, GitHub allows configuring an older image from which to use the layers cache. Official GitHub sources showing the presence of the smells in the platform are included in the replication package \cite{replicationPackage}.

%For three smells, namely {\em Artifacts Fetched By Default}, {\em Use of a Protected Cache}, and {\em No Docker Pull-Through Cache}, we found no evidence of their existence in GitHub Actions. Notably, GitHub presents a similar cache protection mechanism to GitLab; however, unlike GitLab, this mechanism cannot be disabled at the project level. Future work could investigate the presence of additional GitHub-specific smells relevant to practitioners.

\subsubsection*{Practitioners' Discussions}\mbox{}

In the following paragraph, we examine short extracts from two practitioner sources \cite{GitLabPractitioner, GitHubPractitioner} that mention a subset of the smells previously outlined. One source focuses on GitLab CI/CD, while the other on GitHub Actions. For the complete discussions of these and other sources, please refer to our replication package \cite{replicationPackage}. Quotes 1 and 2 present the four extracts. EX1 points at {\em Artifacts Fetched By Default}, while EX2 points at {\em No Docker Layers Cache}. EX3 points both at {\em Artifacts Default Expiration Period} and at {\em Suboptimal Compression Type}, while EX4 at {\em No Docker Pull-Through Cache}. As previously shown, similar problems of missing caches or cache misconfigurations are not exclusive to GitLab, but are also discussed and addressed by practitioners using other CI/CD tools. 

\begin{tcolorbox}[
  colback=white,
  colframe=black,
  boxrule=0.8pt,
  arc=2mm,
  after skip=4pt,
  center title,
  title=\textbf{Learn how to speed up Gitlab CI},
]
\faComment\quad\itshape
EX1: ``Ask yourself this question each time you write a new job: Do I need some files from the cache to run this task? If not, just disable it.''

\faComment\quad\itshape
EX2: ``Each build job is run on a different container with its own Docker daemon. [...] Pull a docker image that you built earlier from your ci image registry before building a new one and using the --cache-from instruction.''
\end{tcolorbox}
\label{Quote:GitLabPractitioner}
\captionof{quoteFloat}{Extracts from a blog post of a digital consulting company \cite{GitLabPractitioner}.}

\begin{tcolorbox}[
  colback=white,
  colframe=black,
  boxrule=0.8pt,
  arc=2mm,
  boxsep=4pt,
  after skip=4pt,
  center title,
  title=\textbf{A Developer's Guide to Speeding Up GitHub Actions},
]
\faComment\quad\itshape
EX3: ``Large or numerous build artifacts and logs can slow down operations. [...] Implement log rotation, compress artifacts, and retain only essential artifacts.''

\faComment\quad\itshape
EX4: ``Consider hosting your own cache server and/or artifact registry that is closer to compute. Example: Uploading large build artifacts or pulling large Docker images from a registry.''
\end{tcolorbox}
\label{Quote:GitHubPractitioner}
\captionof{quoteFloat}{Extracts from a blog post of a CI/CD tooling company \cite{GitHubPractitioner}.}

\subsection{RQ2: Detection Accuracy}
\label{Subsec:RQ2-Results}

To evaluate the accuracy of \ttt{CROSSER}, we built and labeled a dataset ({\em D2}) comprising 82 mature projects, yielding a total of 50 labels per smell. After labeling, we ran \ttt{CROSSER} on {\em D2} to classify the presence of the smells. Table \ref{Tab:RQ2} summarizes the results. 

\ttt{CROSSER} correctly identified the presence of most smell instances, achieving an aggregate $F_{1}$ score of 0.98. Four detectors matched the labels perfectly, while the other three produced six misclassifications (four false positives and two false negatives). We manually reviewed these misclassifications and present the results here. 

In the case of {\em No Dependencies Cache}, one project (\ttt{coolercontrol}) installs libraries using \ttt{apt-get} and caches the entire \ttt{.cache} folder instead of the \ttt{apt-get} folder checked by \ttt{CROSSER}. Another project (\ttt{Remmina}) installs the \ttt{python3-dev} OS package, wrongly identified by \ttt{CROSSER} as a library. Regarding {\em No Docker Layers Cache}, two projects from \ttt{gitlab-org} (\ttt{gitlab} and \ttt{gitlab-runner}), build Docker images without layers cache using commands not detected by \ttt{CROSSER} (e.g., \ttt{docker buildx bake}). Lastly, the detection for {\em No Docker Pull-Through Cache} resulted in two false positives from two \ttt{gitlab-org} projects (\ttt{gitlab} and \ttt{gitlab-foss}). In these two instances, unlike the other misclassifications, the errors are due to the parser, which failed to solve nested variables (\ttt{variables} clause \docscite{S1}) used by some of the workflows' jobs to declare the pulled Docker image. The detector, therefore, could not use the actual image identifier, but only a variable that represented it.

\begin{table}[htbp]
\caption{Summary of RQ2 results: CROSSER's accuracy across the dataset {\em D2}. T = True, F = False, P = Positive, N = Negative.}
\begin{center}
\begin{tabular}{|l|U{0.4cm}U{0.4cm}U{0.4cm}U{0.4cm}|c|}
\hline
\textbf{Detector for Smell} & \textbf{TP} & \textbf{TN} & \textbf{FP} & \textbf{FN} & $\boldsymbol{F_{1}}$\\
\hline
Artifacts Def. Expiration Period & 41 & 9 & 0 & 0 & 1.00\\
Artifacts Fetched By Default & 39 & 11 & 0 & 0 & 1.00\\
No Dependencies Cache & 32 & 16 & 2 & 0 & 0.97\\
No Fallback Cache & 36 & 14 & 0 & 0 & 1.00\\
Redundant Cache Updates & 22 & 28 & 0 & 0 & 1.00\\
No Docker Pull-Through Cache & 43 & 5 & 2 & 0 & 0.98\\
No Docker Layers Cache & 8 & 40 & 0 & 2 & 0.89\\
\hline
\textbf{Total} & 221 & 123 & 4 & 2 & 0.98\\
\hline
\end{tabular}
\end{center}
\label{Tab:RQ2}
\end{table}

These misclassifications highlight the limitations of static analyzers. While most smell instances are correctly identified, non-standard solutions (e.g., niche commands, atypical file paths) and parsing errors affect accuracy. Notably, four of the six errors were found in pipelines configured by the GitLab foundation, which are likely among the most complex on gitlab.com. That is, more complex pipelines are more likely to cause parsing errors or present non-standard commands. Despite these challenges, \ttt{CROSSER} reaches an overall $F_1$ score of 0.98 across {\em D2}. As explained in Section \ref{Subsec:Datasets}, {\em D2} comprises the most mature projects from the larger dataset {\em D1}, further underscoring the robustness of these results. Comparable accuracy has been reported for other static analyzers: Zhang et al.~\cite{zhang2019large} obtained 0.991 recall and 0.998 precision over 15 smells. Vassallo et al.~\cite{vassallo2020configuration} achieved 0.94 recall and 0.87 precision across four smells (reaching perfect accuracy for two of the four smells).

\subsection{RQ3: Smells Frequency}
\label{Subsec:RQ3-Results}

\begin{figure}[htbp]
    \caption{The relative distribution of smelly jobs across the smelly repositories in {\em D1}. The IDs are shown in Section \ref{Sec:Smells}.}
    \centerline{\includegraphics[width=0.5\textwidth]{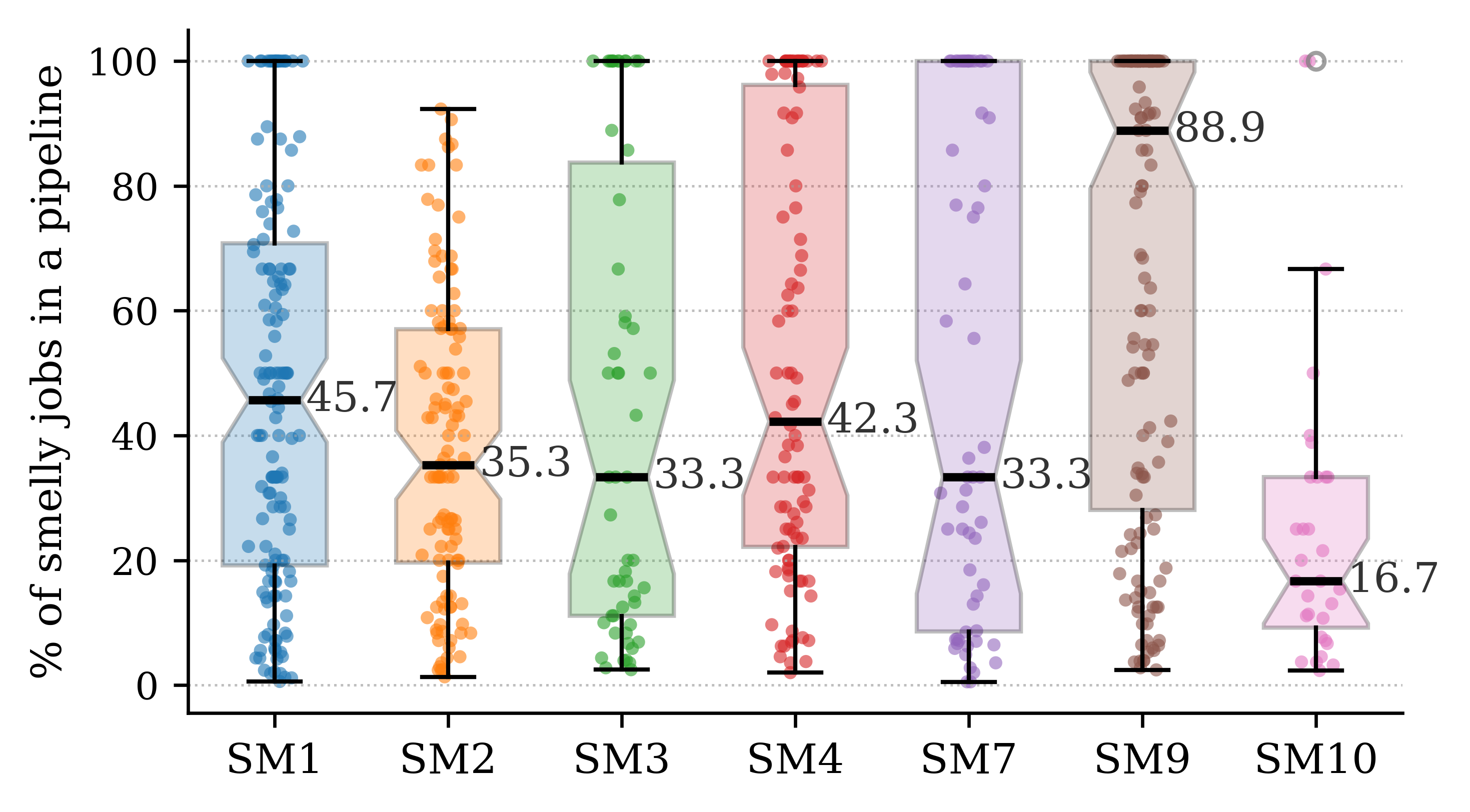}}
    \label{fig:smelly-jobs-distribution}
\end{figure}

In this section, we present the frequency of smells across repositories, by describing three key findings. The data we rely on is summarized in Table \ref{Tab:RQ3-summary} and Fig.~\ref{fig:smelly-jobs-distribution}, obtained running \ttt{CROSSER} on the dataset {\em D1}. Given \ttt{CROSSER}'s accuracy, we expect that the real frequencies of the smells are very close to the ones automatically detected. Additionally, to support our findings, we present examples from the manual inspection of dataset {\em D2}. 

The overall frequencies of smells across {\em D1}, shown in Table \ref{Tab:RQ3-summary}, help us answer RQ3. Nevertheless, it is essential to note that a single smelly job is sufficient to flag the repository as smelly. For this reason, we have processed the distribution of smelly jobs with respect to all jobs in a repository (i.e., if one job is smelly and the pipeline has ten jobs, then the workflow has 10\% of smelly jobs). Fig.~\ref{fig:smelly-jobs-distribution} depicts the result. 

%Although these statistics help us answer RQ3 in more detail, they are still influenced by {\em irrelevant} jobs, i.e., jobs that cannot contain a particular smell, because they do not present the conditions for the smell to happen, for instance, they do not define the \ttt{artifacts} keyword. We, therefore, define the concept of {\em Potentially Smelly Job} (PSJ) based on the descriptions from Section \ref{Sec:Smells}. The distribution of PSJs over SJs is shown in Table \ref{Tab:RQ3-smell-risk-per-job}. 

In the following paragraphs, we provide three empirical findings about the frequency of smells. Whenever applicable, we discuss similar results from related work.

\begin{table}[htbp]
\caption{Summary of RQ3 results: the smells frequency in the dataset {\em D1}. $^{\dag}$ = pipelines using Python. $^*$ = group repositories.}
\begin{center}
\begin{tabular}{|l|U{1.5cm}|U{1.5cm}|}
\hline
\multirow{2}{*}{\textbf{Smell}} 
& \multicolumn{2}{c|}{\textbf{Repositories Number}} \\
\cline{2-3}
& \textbf{Smelly} & \textbf{Applicable} \\
\hline
Artifacts Default Expiration Period & 144 (63.2\%) & 228 \\
Artifacts Fetched By Default & 114 (50.0\%) & 228 \\
No Dependencies Cache & 54 (68.4\%) & 79$^{\dag}$ \\
No Fallback Cache & 96 (42.1\%) & 228 \\
Redundant Cache Updates & 59 (25.9\%) & 228 \\
No Docker Pull-Through Cache & 146 (85.9\%) & 170$^*$ \\
No Docker Layers Cache & 31 (13.6\%) & 228 \\
\hline
\multirow{2}{*}{\textbf{Property Affected}} 
& \multicolumn{2}{c|}{\textbf{Repositories Number}} \\
\cline{2-3}
& \textbf{Smelly} & \textbf{Applicable} \\
\hline
Speed & 190 (83.3\%) & 228 \\
Efficiency & 203 (89.0\%) & 228 \\
Reliability & 202 (88.6\%) & 228 \\
\hline
\end{tabular}
\label{Tab:RQ3-summary}
\end{center}
\end{table}

\subsubsection*{\textbf{FINDING 1: Cache-related smells are very widespread in mature open-source projects}}
We have detected that only 11.0\% of the repositories are smell-free (25 out of 228), and the median number of smells per project is 3. Furthermore, all properties (speed, efficiency, and reliability) are impacted in over 80\% of CI/CD pipelines. When it comes to singular smells, the most common is {\em No Docker Pull-Through}, but three more smells are present in at least half of the applicable repositories. Lastly, when focusing on smelly projects, the smells we detect affect many jobs, not just a few. For example, as shown in Fig.~\ref{fig:smelly-jobs-distribution}, {\em No Docker Pull-Through Cache} is found in 88.9\% (median) of jobs. All other smells affect between one-sixth and half of pipeline jobs (median values). These observations are consistent with previous reports of CI/CD smells and anti-patterns. Zhang et al.~\cite{zhang2022buildsonic} found that 99\% of the projects in their dataset exhibit at least one of 15 smells, with an average of five smells per project. Vassallo et al.~\cite{vassallo2019automated} reported that three of the four detected smells are present in almost all of the projects they investigated, while the remaining smell was found in 15 projects (41.7\%).

%\subsubsection{\textbf{Whenever a smell can potentially occur, it will likely occur}}
%When we consider jobs potentially smelly, 63.0\% of all jobs actually become smelly, as shown in Table \ref{Tab:RQ3-smell-risk-per-job}. This ratio varies from smell to smell, ranging from 96.0\% for {\em No Fallback Cache} to 26.2\% for {\em No Docker Pull-Through Cache}. 

%Notably, we can see that almost 40\% of Docker images are built with a layer cache, while only 15\% of installed libraries are cached between pipeline runs. This is likely because many practitioners do not believe that library caches result in noticeable performance and reliability differences, despite empirical evidence indicating otherwise.

\subsubsection*{\textbf{FINDING 2: Developers often leave default values}}
Predictably, developers leave default values in their configuration files. As shown in Table \ref{Tab:RQ3-summary}, {\em Artifacts Default Expiration Period}, and {\em Artifacts Fetched By Default} are present in at least half of the repositories. Although default behaviour is designed for convenience, it ultimately results in slower, less efficient, and less reliable pipelines. Similarly, Zhang et al.~\cite{zhang2022buildsonic} note that their {\em performance-related configuration smells} are often introduced by default configurations.

\subsubsection*{\textbf{FINDING 3: Developers are potentially not aware of advanced CI/CD cache-related features}}
85.9\% of all group repositories in {\em D1} is affected by {\em No Docker Pull-Through Cache}. According to the release notes \docscite{S3}, the feature enabling the pull-through cache was extended to free-tier users in December 2020, i.e., more than four years before our mining process took place. Nevertheless, the widespread lack of container registries is significant, especially since developers can simply add a prefix to each Docker Hub image to use the feature. Similarly, we noticed that only four repositories in {\em D1} used a fallback cache at least once, even though the feature came out in GitLab 16.0 \docscite{S13} (i.e., May 2023).

Given these considerations, we expanded the investigation to two additional features associated with two smells from Section \ref{Subsec:Smells-Catalog-Method}: multiple caches per job (from {\em God Cache}) and custom compression types (from {\em Suboptimal Compression Types}). We have discovered that only 5.3\% of repositories used multiple caches per job at least once, and 3.1\% used a custom compression type at least once. Note that the lack of these features is not enough to determine the presence of the smells, which, as outlined in Section \ref{Subsec:Smells-Catalog-Method}, are complex to detect automatically. Rather, it only confirms the absence of advanced cache-related features in mature open-source repositories.

\section{Discussion}
\label{Sec:Discussion}

In this work, we have identified ten cache-related smells that affect speed, efficiency, and reliability and are relevant to practitioners. Although the GitLab documentation includes webpages dedicated to practical tips for caching \docscite{S6, S10}, we still found additional best practices and smells scattered throughout other webpages. It is therefore challenging for practitioners to learn and internalize all the relevant smells to avoid, underscoring the need to mine a unified catalog. CI/CD platforms should review their documentation to be more exhaustive, especially when presenting guidelines. In addition, CI/CD platforms should include quantitative examples for the best practices and smells they discuss. In fact, during our mining process, we did not find measures such as ``Fetching default dependencies delays a job by at least 10 seconds''.

To automatically detect seven of the ten smells, we have developed \ttt{CROSSER}, a static analyzer that achieves very high accuracy ($F_1$ score of 0.98) across 82 mature open-source projects. When labeling {\em D2}, we found that manually identifying smells from CI/CD configurations is time-consuming. Mature projects often rely on higher-level GitLab CI/CD features, such as inherited global values and variables (see Section \ref{Subsec:CROSSER}), which increase conciseness but reduce readability and make analysis difficult. We believe that CI/CD platforms should be more transparent about the values of jobs during pipeline runs, so smells can be reduced. Using as an example one repository from {\em D1}, \ttt{cable-mc/cobblemon} sets the artifacts' expiration date to 7 days. Nevertheless, since this is done using the \ttt{default} keyword rather than YAML anchors, the value is not considered by the jobs, so the actual retention period is 30 days. 

Lastly, we have shown the widespread presence of smells, affecting nearly 8 out of 9 mature open-source repositories. This underscores the need for both a smells catalog and an effective detection tool. We believe that, especially for smells that are easy to detect (e.g., {\em Artifacts Default Expiration Period}, {\em No Fallback Cache}), CI/CD platforms should integrate automated checks. In addition, while we have proposed a heuristics-based tool for detection, future work could analyze the accuracy of pre-trained LLMs, which remove the need to develop parsers. This is even more relevant for the two smells for which we did not define any heuristic (i.e., {\em God Cache}, {\em Suboptimal Compression Type}). Such smells may be better suited to LLMs or, potentially, to a combination of tool-based static extraction for the relevant smell context and LLM querying only on that context. This combination may be necessary, as many CI/CD configuration files from {\em D1} contain thousands of lines of code.

\section{Threats to Validity}
\label{Sec:Threats}

Regarding {\bf construct validity}, our study is threatened by the lack of quantitative validation of our smells catalog and by the use of blog-like sources to assess the relevance of the smells to practitioners. The former threat aligns with prior works in the CI/CD community that propose patterns and smells \cite{yin2024developeraccelerationsci, vassallo2019automated, vassallo2020configuration, zampetti2020empirical, shahin2017continuous, gallaba2018use, sharma2016does}. To counter the threat, we derived our catalog from GitLab sources as done in previous works \cite{zhang2022buildsonic, gallaba2018use, sharma2016does, garousi2019guidelines}. Following official sources enhances the credibility and relevance for real-world usage. Additionally, we relied on existing literature that measured the effects of caches on speed, efficiency, and reliability in CI/CD systems \cite{zhang2022buildsonic, gallaba2020accelerating, ghaleb2022studying}. Notably, Ghaleb et al.~\cite{ghaleb2019empirical} discovered that misconfigured caches do not bring the expected improvements. The latter threat, i.e., validation against a pre-existing corpus, may lead to the exclusion of relevant smells. Although other studies \cite{zampetti2020empirical,vassallo2019automated} used developer surveys to validate their catalogs, Garousi et al.\ \cite{garousi2019guidelines} highlight the importance of the grey literature for understanding practitioners' concerns and interests. In our work, we used a corpus focused on the performance and reliability of CI/CD pipelines and encompassing multiple CI/CD technologies (e.g., GitLab CI/CD, GitHub Actions), improving the generalizability. Moreover, our corpus size is comparable to other studies using blog-like sources \cite{urdih2025architectural,warnett2022architectural,ntentos2024supporting}.

Regarding {\bf internal validity}, our coding and labeling may be subject to human error and bias. To minimize this threat, two authors coded the grey sources independently and compared their notes at the end. To minimize the labeling risk, we have adopted a systematic approach and cross-checked ambiguous cases between all authors. Nevertheless, misclassifications or oversights may still have occurred, particularly in borderline cases.

Finally, a key threat to {\bf external validity} is the generalization of our results on accuracy and frequency. Our project selection process may bias our findings. We mitigated the risk by using inclusion/exclusion criteria that reduce toy projects \cite{munaiah2017curating}, without further filtering the projects' characteristics (e.g., size). Notably, our datasets include a wide range of programming languages and project types. To study \ttt{CROSSER}'s accuracy, we ranked projects based on five maturity metrics proposed by Munaiah et al.\ \cite{munaiah2017curating}. Since mature projects use more complex CI/CD pipelines, we believe that \ttt{CROSSER}'s measured accuracy reflects its real-world performance.

\begin{comment}
## Artifacts Default Expiration Period

GitHub equivalent:
- https://docs.github.com/en/repositories/managing-your-repositorys-settings-and-features/enabling-features-for-your-repository/managing-github-actions-settings-for-a-repository#configuring-the-retention-period-for-github-actions-artifacts-and-logs-in-your-repository
- https://github.com/actions/upload-artifact?tab=readme-ov-file#inputs
- https://docs.github.com/en/actions/how-tos/manage-workflow-runs/remove-workflow-artifacts#setting-the-retention-period-for-an-artifact

## Use of a Protected Cache

GitHub equivalent:
- https://docs.github.com/en/actions/reference/workflows-and-actions/dependency-caching#restrictions-for-accessing-a-cache

## Suboptimal Compression Type

GitHub equivalent:
- https://github.com/actions/upload-artifact?tab=readme-ov-file#altering-compressions-level-speed-v-size

## No Docker Layers Cache

GitHub equivalent:
- https://github.com/docker/build-push-action?tab=readme-ov-file#inputs
- https://docs.docker.com/build/ci/github-actions/cache/

\end{comment}

\section{Related Work}
\label{Sec:Related-Work}

In this section, we discuss how our work compares with existing CI/CD studies, focusing on other catalogs, caching, and detectors.

\subsection{CI/CD Patterns and Smells Catalogs}

Numerous works have proposed catalogs of CI/CD patterns \cite{duvall2007continuous, duvall2010continuous, duvall2011continuous, dullmann2021stalkcd, shahin2017continuous}, as well as their violation in the form of smells \cite{vassallo2019automated, vassallo2020configuration, zhang2022buildsonic, gallaba2018use, zampetti2020empirical, sharma2016does}. Yin et al.\ \cite{yin2024developeraccelerationsci} inspected 2896 jobs to identify how developers improve the performance, resulting in 16 patterns for speed and efficiency. Zhang et al.\ \cite{zhang2022buildsonic} proposed a catalog of 20 performance-related smells mined from the documentation of \ttt{TravisCI}, \ttt{Maven}, and \ttt{Gradle}. Gallaba et al.\ \cite{gallaba2018use} studied \ttt{TravisCI}'s sources and came up with four CI misconfigurations. These works have a broader scope than ours, with catalogs not focused on caching. Unlike other studies \cite{vassallo2020configuration, gallaba2018use, zhang2022buildsonic,yin2024developeraccelerationsci}, we investigated the presence of our catalog in another CI/CD technology to enhance its relevance.

To derive patterns or smells, some studies used a scientific literature review \cite{duvall2007continuous, shahin2017continuous}, while others, like ours, used grey literature sources \cite{zampetti2020empirical, zhang2022buildsonic, gallaba2018use, sharma2016does, warnett2022architectural}. Similar to Zhang et al.\ \cite{zhang2022buildsonic}, Gallaba et al.\ \cite{gallaba2018use}, and Sharma et al.\ \cite{sharma2016does}, we mined only the official tools' sources to improve their authority \cite{garousi2019guidelines}. After the mining process, we used practitioners' views to understand the interest in the candidate smells. Similarly, Urdih et al.~\cite{urdih2025architectural} and Warnett et al.~\cite{warnett2022architectural} proposed patterns using, respectively, 38 and 35 blog-like sources. Other researchers \cite{vassallo2019automated,zampetti2020empirical} used developer surveys instead. 

\subsection{Caching in CI/CD}

Caching is a best practice for many kinds of systems, and CI/CD workflows are no exception. Ghaleb et al.~\cite{ghaleb2019empirical} studied the pipelines of 67 GitHub projects and discovered a strong association between caching and faster pipelines. In addition, they discovered that misconfigured caches were often slower. In a subsequent study \cite{ghaleb2022studying}, they investigated a corpus of 588 projects and 924,616 pipelines and showed that caches are the most effective way to both speed up pipelines and reduce the risk of failure. In another study, Bouzenia et al. \cite{bouzenia2024resource} analyzed 1.3 million workflow runs from 952 mature repositories and found that projects using dependency caching saved, on average, 4\% of pipeline runtime.

Other works proposed automatic tools to detect and repair missing caches in CI/CD workflows. Celik et al.~\cite{celik2016build} presented \ttt{MOLLY}, a system that automatically saves the necessary Java libraries. Zhang et al.\ \cite{zhang2022buildsonic} developed \ttt{BuildSonic} that detects and repairs missing library caches for Java applications in \ttt{TravisCI} \cite{travisCI}. Gallaba et al.\ \cite{gallaba2020accelerating} proposed \ttt{KOTINOS}, which automatically caches pipeline dependencies by monitoring system calls. Esfahani et al.~\cite{esfahani2016cloudbuild} presented \ttt{CloudBuild}, a distributed caching system used by Microsoft that saves library dependencies by extracting them from the build files. Each of these papers observed improved pipeline performance across most of their subject systems when applying the cache.

Unlike our work, these studies focus only on dependency caches and overlook other types of caches, except for \ttt{KOTINOS} \cite{gallaba2020accelerating}, which caches Docker layers. Furthermore, they do not consider cache misconfigurations that affect performance and reliability, nor do they consider any smells related to reliability. Lastly, these works fail to address the features (e.g., fallback caches) and pitfalls (e.g., empty Docker Daemon) that CI/CD platforms present for caching.

\subsection{Smells Detectors in CI/CD}

Several works proposed tools to detect CI/CD smells. Vassallo et al.\ \cite{vassallo2019automated} developed \ttt{CI-ODOR}, a tool to identify four smells based on build logs. In another study, Vassallo et al.\ \cite{vassallo2020configuration} presented \ttt{CD-LINTER}, which identifies four smells relying on the parsed representation of GitLab CI/CD files. This approach, i.e., detecting smells only by statically analyzing CI/CD files, has proved successful and was also applied by Zhang et al.\ \cite{zhang2022buildsonic} for 15 performance-related smells. Similarly, Gallaba et al.\ \cite{gallaba2018use} detect four misconfigurations in Travis CI. Like in our case, all these researchers showed very high accuracy metrics. Gallaba et al.\ \cite{gallaba2018use} reported a recall of 0.828. Vassallo et al.\ \cite{vassallo2020configuration} achieved  0.94 recall and 0.87 precision (reaching 1.00 for two of the four smells). Zhang et al.\ \cite{zhang2022buildsonic} obtained 0.991 recall and 0.998 precision. Notably, they reached a perfect (1.0) $F_{1}$ score for their smell {\em No Dependencies Cache}, while \ttt{CROSSER} achieved 0.97. 

\section{Conclusions and Future Work}
\label{Sec:Conclusions}

Prior research has demonstrated the importance of caches for the performance and reliability of CI/CD pipelines \cite{ghaleb2022studying, gallaba2020accelerating, zhang2022buildsonic, ghaleb2022studying}. However, existing works have focused entirely on dependency caches, ignoring other types of caches and cache misconfigurations. In this paper, we systematically evaluated the official GitLab CI/CD documentation to create a catalog of ten cache-related smells, which negatively impact speed, efficiency, and reliability.

To address these issues, we introduced \ttt{CROSSER}, a static analysis tool that can detect seven of the ten identified smells. \ttt{CROSSER} achieves a high level of accuracy with an overall $F_{1}$ score of 0.98 across 82 mature projects, matching or even exceeding the performance of existing similar tools~\cite{vassallo2020configuration, gallaba2018use, zhang2022buildsonic}. Despite the accuracy reached by static analyzers, the effort required to develop them is still high, and requires an in-depth knowledge. Future work could, therefore, analyze the accuracy of pre-trained LLMs. 

When applying our catalog and tool to a dataset of 228 mature repositories, we found that cache-related smells are widespread: only 11\% of projects were smell-free, with a median of three smells per repository. Notably, the most common smell, {\em No Docker Pull-Through Cache}, affects 85.9\% of group projects. These results suggest that many practitioners may not be aware of the full scope of caching capabilities available in GitLab, leading to suboptimal CI/CD workflows. This highlights the need for smell catalogs summarizing bad practices, and automated tools that assist in the detection of these bad practices, such as \ttt{CROSSER}. Future work includes extending the catalog and detection tool to other CI/CD platforms, as well as developing automated repair approaches using either static analysis tools or LLMs to quantify the impact of cache-related smells on pipeline performance and reliability.

\begin{acks}
We acknowledge the use of GitHub Copilot (with model GPT-4o) to assist in the development of \ttt{CROSSER}.

This research was funded in whole or in part by the Austrian Science Fund (FWF) project CQ4CD, Grant-DOI: 10.55776/I6510, and by the FFG (Austrian Research Promotion Agency) project MODIS (no.\ FO999895431) and project BEAM (no.\ FO999933314).
\end{acks}

\bibliographystyle{ACM-Reference-Format}
\bibliography{references}

\end{document}